\documentclass[twocolumn,superscriptaddress,pra,longbibliography,showpacs,preprintnumbers,amsmath,amssymb,floatfix]{revtex4-1}
\usepackage{titlesec}
\usepackage{amsmath}
\usepackage{bm}
\usepackage{graphicx}
\usepackage[ansinew]{inputenc}
\usepackage{array}
\usepackage{color}
\usepackage{amsxtra}
\usepackage{amstext}
\usepackage{amssymb}
\usepackage{latexsym}
\usepackage{gensymb}
\usepackage{dsfont}
\usepackage{braket}
\usepackage[caption=false]{subfig}
\usepackage[makeroom]{cancel}

\usepackage{amsmath,amscd}

\usepackage{balance}

\usepackage{dcolumn}
\usepackage{bm}
\usepackage{bbm}
\usepackage{braket}
\usepackage{mathrsfs}
\usepackage{amstext}
\usepackage{euscript}
\usepackage{txfonts}

\usepackage[dvipsnames]{xcolor}

\usepackage[unicode=true,
 bookmarks=false,
 breaklinks=false,pdfborder={0 0 1},backref=false,colorlinks=true,citecolor=blue]
 {hyperref}

\usepackage{tikz-cd}

\makeatletter
\@ifundefined{textcolor}{}
{%
 \definecolor{BLACK}{gray}{0}
 \definecolor{WHITE}{gray}{1}
 \definecolor{RED}{rgb}{1,0,0}
 \definecolor{GREEN}{rgb}{0,1,0}
 \definecolor{BLUE}{rgb}{0,0,1}
 \definecolor{CYAN}{cmyk}{1,0,0,0}
 \definecolor{MAGENTA}{cmyk}{0,1,0,0}
 \definecolor{YELLOW}{cmyk}{0,0,1,0}
}

\makeatletter
\renewcommand*\env@matrix[1][*\c@MaxMatrixCols c]{%
  \hskip -\arraycolsep
  \let\@ifnextchar\new@ifnextchar
  \array{#1}}
\makeatother

\newcommand{\cref}[1]{Ref.\,\cite{#1}}

\makeatother

\begin{document}

\title{Quantum Speed Limits and  the Ultimate Scaling of the Quantum Sensors}

\author{Yusef Maleki}
\email{maleki@physics.tamu.edu}
\affiliation{Department of Physics and Astronomy, Texas A\&M University, 
	College Station, Texas 77843-4242}

\date{\today}

\begin{abstract}

Quantum metrology promises sensitivity beyond classical strategies, yet it remains unsettled how quantum-enabled precision should scale with physical resources and how to interpret quantum advantage. We provide a physically grounded resource accounting that clarifies the true Heisenberg limit and resolves apparent super-Heisenberg paradoxes. We demonstrate that the Heisenberg limit is best viewed as an information-theoretic manifestation of the quantum speed limit. We illustrate these ideas with a simple, super-resolving phase-estimation protocol based on Rabi oscillations in two-level atoms driven on an $m$-photon resonance. In this setting, the phase error scales as $n^{-m/2}$, where $n$ is the average photon number. Recasting metrological sensitivity through quantum dynamical speed limits yields operational bounds that reconcile such super-resolution strategies with the standard Heisenberg interpretation and identify the relevant resources in the norm of the generator. We also revisit the common attribution of the NOON state's $1/n$ scaling to quantum entanglement. We show that such an attribution is not generic and the Heisenberg $1/n$ scaling does not, by itself, certify entanglement as the enabling resource.

\end{abstract}

\maketitle

\section{ Introduction}

Advances in precision measurements have transformed our understanding of nature, enabling technologies that range from gravitational wave observatories~\cite{tse2019quantum,jia2024squeezing} to super-resolved biological imaging~\cite{taylor2016quantum,khashami2023fundamentals,couteau2023applications}. At the core of these capabilities lies the ability to estimate unknown parameters from experimental data using optimally designed measurement protocols. The fundamental precision limit in such estimation tasks is dictated by the Cram\'er--Rao bound, which relates the minimum variance of any unbiased estimator to the Fisher information~\cite{Giovannetti,maleki2022distributed}. In a foundational contribution, Helstrom extended this concept to quantum systems by formulating an estimation bound for parameters embedded in a quantum state~\cite{helstrom1967minimum}. Later, Caves demonstrated that quantum strategies can offer enhanced sensitivity compared to classical methods, assuming a fixed amount of physical resources~\cite{caves1981quantum}.

Over time, quantum sensing has progressed from a primarily foundational subject into a rapidly expanding area of quantum science and technology, with increasingly practical implementations \cite{yu2021molecular,aslam2023quantum,khashami2025quantum}. At the same time, the field has moved beyond ideal unitary parameter-estimation models toward more realistic scenarios that incorporate environmental noise, many-body correlations, and platform-dependent experimental constraints. Several review articles have discussed major developments in quantum metrology and sensing, spanning basic theory, experimental implementation, and emerging applications \cite{degen2017quantum,pirandola2018advances,braun2018quantum,demkowicz2020multi,albarelli2020perspective,liu2020quantum,sidhu2020geometric,huang2024entanglement}. Most of these reviews focus on specific aspects of the field, including multiparameter estimation \cite{demkowicz2020multi,albarelli2020perspective,liu2020quantum}, estimation in quantum many-body systems \cite{montenegro2408review}, parameter estimation in continuous-variable systems \cite{fadel2025quantum}, geometric approaches \cite{sidhu2020geometric}, and applications in areas such as particle physics \cite{bass2024quantum} and NISQ-era quantum devices \cite{meyer2021fisher}.

A key distinction between classical and quantum estimation strategies lies in how the estimation error scales with the number of resources. In classical protocols using $n$ independent probes, the Fisher information can scale linearly with $n$ at most, producing an estimation error that decreases as $1/\sqrt{n}$. This scaling defines the shot-noise limit (SNL), also known as the standard quantum \mbox{limit~\cite{feng2014super,maleki2021quantum}.} In contrast, quantum-enhanced strategies can exploit non-classical resources such as  entanglement to achieve a quadratic scaling of the Fisher information, enabling an error that scales as $1/n$, the so-called Heisenberg limit (HL). This ultimate bound in precision can be attained using maximally path-entangled states of the form \mbox{$(|n,0\rangle + |0,n\rangle)/\sqrt{2}$~\cite{Xiang,maleki2021quantum}.} The ability to surpass the SNL and approach the HL forms the basis for emerging quantum metrological technologies, offering unprecedented sensitivity and resolution in measurement tasks~\cite{zhou2018achieving,tse2019quantum,taylor2016quantum}.

One prominent illustration of quantum-enhanced resolution is found in interferometric lithography, where spatial features can be resolved beyond the classical diffraction limit, defined by the Rayleigh criterion~\cite{Boto}. This enhancement relies on photon-number entangled states, highly nonclassical quantum states that are both difficult to generate and extremely sensitive to decoherence~\cite{maleki2018generating,pezze2018quantum}. To address these practical limitations, alternative schemes have been developed that exploit classical light pulses in conjunction with multiphoton resonance effects~\cite{hemmer2006quantum,liao2010quantum,wolk2012phase}. Notably, it has been demonstrated that a carefully engineered narrow-band multiphoton detection process can enable subwavelength resolution even in the absence of entanglement and with classical strategies, offering a robust and experimentally accessible route to quantum-inspired lithographic precision~\cite{hemmer2006quantum}.

Furthermore, recent theory and experiments indicate that classical-nonlinear and interaction-based strategies can surpass the standard Heisenberg scaling. Beltr{\'a}n and Luis predicted an $n^{-3/2}$ phase-error law for systems with classical optical \mbox{nonlinearities \cite{beltran2005breaking},} later observed experimentally by Napolitano \emph{et al.} \cite{napolitano2011interaction}. More generally, Boixo \emph{et al.} reported scalings of the form $n^{-q}$ with integer $q>0$ in interaction-driven \mbox{metrology \cite{boixo2007generalized}.} 
Such scaling beyond $1/n$ is sometimes referred to as super-Heisenberg \mbox{behaviors \cite{napolitano2011interaction,rams2018limits}} and shows that the familiar $1/n$ limit is not a universal ceiling, and its applicability depends on how resources are counted and which Hamiltonian generators are allowed. This motivates a central question: which quantum-mechanical principle sets the ultimate scaling of parameter-estimation precision? And in what sense is $1/n$ scaling considered ultimate?

Considering this line of thought, the connection between Fisher information and quantum dynamics has also provided a deeper geometric interpretation of precision limits. In quantum estimation theory, the quantum Fisher information quantifies the infinitesimal distinguishability of neighboring quantum states and defines the statistical distance associated with the Bures metric \cite{taddei2013quantum}. From this perspective, a larger quantum Fisher information means that the probe state moves faster through Hilbert space under a change of the estimated parameter, making nearby parameter values more distinguishable. This geometric viewpoint naturally links metrological sensitivity to quantum speed limits, which bound the minimum time required for a quantum system to evolve between distinguishable states \cite{maleki2023speed}.

Despite these developments, the relation between the Heisenberg limit \cite{carrasco2026time,holland1993interferometric,maleki2018recovery}, super-Heisenberg scaling \cite{napolitano2011interaction,rams2018limits}, and quantum speed limits \cite{ness2021observing,chen2023speed,maleki2020speed} remains conceptually subtle, especially in nonlinear or interaction-based metrological protocols. In such systems, apparent improvements beyond the conventional $1/n$ scaling may arise because the relevant generator is no longer linear in the number of probes or photons. Therefore, a meaningful comparison of different strategies requires careful identification of the physical resource that drives distinguishability. This motivates the present work, where we use quantum speed limits as an operational framework to clarify the ultimate scaling of quantum sensors and to distinguish genuine precision enhancement from changes in resource accounting.

In this work, we explore a minimal yet powerful framework for superresolving parameter estimation, based on the Rabi dynamics of two-level atomic systems. These atoms undergo Rabi oscillations triggered by the simultaneous absorption of $m$ photons. At the heart of this protocol lies the phenomenon of multiphoton resonance, wherein excitation occurs only when the atom absorbs exactly $m$ photons from the field. Precision in time estimation can be achieved by monitoring observables such as population inversion or the full Rabi cycle of the atomic system. We show that when the average photon number in the driving field is $n$, the achievable precision in estimating time can scale as $n^{-m/2}$, which seemingly surpasses the  so-called Heisenberg limit $1/n$, for $m>2$.
To shed light on the basis of the ultimate scaling, we show that the notion of quantum speed limit lends useful tools. In other words,
to uncover the fundamental limits of this scaling, we formulate the estimation sensitivity within the framework of quantum dynamical speed limits. This approach offers a deeper understanding of the system's metrological potential and sheds light on the apparent paradoxical features between superresolution strategies and the conventional interpretation of the Heisenberg limit. 

Furthermore, we revisit a common view that the $1/n$ Heisenberg scaling of NOON states is a direct consequence of mode entanglement. We  give an example demonstrating that this association is not necessarily valid or generic. In fact, we demonstrate that the scaling need not originate from entanglement. In particular, we demonstrate that an experimentally accessible single-mode probe can attain the same Heisenberg-limited $1/n$ sensitivity without invoking mode entanglement.

\section{ Quantum Speed Limits and Fisher Information: A Dynamical Perspective on the Metrological Scaling}

To ground the discussion in a familiar metrological setting, we begin with a canonical probe that attains the Heisenberg 
$1/n$  scaling in phase estimation. This choice provides a clear benchmark against which the dynamical bounds derived later can be compared. To this end, we take the maximally path-entangled NOON state as our reference probe and consider its evolution under a phase shift generated by a fixed Hamiltonian. The resulting sensitivity will shed light on our metrological resource accounting and speed-limit analysis. We therefore will use it to connect phase sensitivity with distinguishability in the Bures geometry and with the quantum Fisher information. 
Interpreting phase sensitivity as geodesic speed in the Bures metric makes the link to the quantum Fisher information explicit and lets us import quantum speed limits as metrological bounds, not merely a dynamical constraint.
So, we begin with the NOON state, defined as
\begin{equation}
|\psi \rangle = \frac{1}{\sqrt{2}} \left( |n,0\rangle + |0,n\rangle \right),
\end{equation}
which undergoes a phase shift $\varphi$ implemented by the unitary operator
$U(\varphi) = e^{-i\varphi a^{\dagger}_2 a_2}$,
where $a$ and $a^{\dagger}$ are the photon annihilation and creation operators. Under this transformation, the state evolves to
\begin{equation}
|\psi(\varphi) \rangle = \frac{1}{\sqrt{2}} \left( |n,0\rangle + e^{-i \varphi n} |0,n\rangle \right),
\end{equation}
leading to a phase sensitivity that scales as $\delta \varphi \propto 1/n$, the Heisenberg limit~\cite{zwierz2010general}. In this strategy, entanglement plays a central role in surpassing the shot-noise limit, which constrains precision to $\delta \varphi \propto 1/\sqrt{\bar{n}}$. The  accumulation of phase using $n$ entangled photons enables enhanced sensitivity, showing the quantum origin of the improved scaling.
The phase shift operator is defined as $U(\varphi) = e^{-i\varphi a^{\dagger}_2 a_2}$,
which arises from the Hamiltonian $H = \hbar \omega a^{\dagger}_2 a_2,$
generating time evolution through $U(t) = e^{-i\omega t a^{\dagger}_2 a_2}$, where $t$ is the duration of evolution and $\varphi = \omega t$. Hence, estimating $\varphi$ is equivalent to estimating the time parameter $t$ for a fixed Hamiltonian. 

To illustrate the connection between the sensitivity and state geometry, we consider the NOON state undergoing a short-time evolution
\begin{equation}
\vert \psi(\delta t) \rangle = \frac{1}{\sqrt{2}} \left( |n,0\rangle + e^{-i n \omega \delta t} |0,n\rangle \right).
\end{equation}

This yields the transition probability
\begin{equation}
\left| \langle \psi(0) | \psi(\delta t) \rangle \right|^2 = \cos^2\left(\frac{n \omega \delta t}{2}\right),
\end{equation}
leading to a time sensitivity that scales as $\delta t \propto 1/n$. Therefore, the scaling of the phase estimation is fundamentally connected to the distinguishability of the initial and final states of the quantum probes.

Alternatively, for a coherent state~\cite{scully1999quantum}
\begin{equation}
\vert \alpha \rangle = e^{-|\alpha|^2/2} \sum_{n=0}^\infty \frac{\alpha^n}{\sqrt{n!}} \vert n \rangle,
\end{equation}
the evolution under $U(t)$ results in a displaced coherent state $\vert e^{-i\omega \delta t} \alpha \rangle$, with the overlap
\begin{equation}
|\langle \alpha | \alpha e^{-i\omega \delta t} \rangle|^2 = \exp\left[-|\alpha(1 - e^{i\omega \delta t})|^2\right] \approx \exp\left[-\langle n \rangle (\omega \delta t)^2\right],
\end{equation}
where $\langle n \rangle = |\alpha|^2$. This gives the expected shot-noise scaling $\delta t \propto 1/\sqrt{\langle n \rangle}$.

These examples demonstrate that phase estimation is deeply connected to the time required for the system to evolve into a distinguishable state. This connection can be formalized more generally using the quantum Cram\'er--Rao bound, which constrains the precision of time estimation for an unbiased estimator as~\cite{maleki2022distributed,maleki2021quantum1}
\begin{equation}\label{CRB}
\delta t \geq \delta t_{\mathrm{CRB}} = \frac{1}{\sqrt{\mathcal{F}_Q(t)}},
\end{equation}
where $\mathcal{F}Q(t) = \mathrm{Tr}[\rho(t) L(t)^2]$ is the quantum Fisher information (QFI)~\cite{Giovannetti,maleki2022distributed}. The symmetric logarithmic derivative $L(t)$ is defined implicitly via
\begin{equation}
\partial_t \rho(t) = \frac{1}{2} \left[ \rho(t) L(t) + L(t) \rho(t) \right],
\end{equation}
where $\rho(t)$ is the system time-dependent density matrix. The distance between two density matrices can be quantified using the Bures angle~\cite{bures1969extension}, as
\begin{equation}
\mathcal{L}(\rho(0), \rho(t)) = \arccos\left( \sqrt{F(\rho(0), \rho(t))} \right),
\end{equation}
where $F(\rho(0), \rho(t))$ denotes the fidelity between the two density matrices $\rho(0)$ and $\rho(t)$~\cite{uhlmann1976transition, jozsa1994fidelity}, given by
\begin{equation}
F(\rho(0), \rho(t)) = \left[ \mathrm{Tr} {\sqrt{ \sqrt{\rho(0)} \rho(t) \sqrt{\rho(0)} }} \right]^2.
\end{equation}

For infinitesimal changes in time, the Bures angle reduces to the Bures metric, expressed as
\begin{equation}
dS_{B}^{2} = 2 - 2 \sqrt{F(\rho(t), \rho(t + dt))}.
\end{equation}

For a normalized pure state $|\Psi(t)\rangle$, the quantum Fisher information (QFI) for the time parameter can be written as \cite{,maleki2024universal}
\begin{equation}\label{Fisher}
\mathcal{F}_Q(t)=4\Big(\langle \partial_t\Psi(t)|\partial_t\Psi(t)\rangle-|\langle \Psi(t)|\partial_t\Psi(t)\rangle|^2\Big).
\end{equation}

On the other hand, there is an important  relation between the fidelity and the QFI, $\mathcal{F}_Q(t)$ \cite{taddei2013quantum}
\begin{align}
   F[\rho(t), \rho(t+d t)]=1-\frac{\mathcal{F}_{Q}(t)}{4} d t^{2}+\mathcal{O}(d t^{4}).
\end{align}

This interesting relation shows that Fisher information measures the rate at which two states become distinguishable as a specific parameter changes. Consequently, the speed of the evolution of the system is determined by \cite{pezze2018quantum,maleki2024universal}
\begin{equation}
\mathcal{V}(t)=
\frac{d S_{B}}{dt}=\frac{1}{2}\sqrt{\mathcal{F}_{Q}(t)}.   
\end{equation}

This deep connection between quantum speed limit (QSL) \cite{taddei2013quantum,deffner2013quantum} and Fisher information shows that the rate of dynamics at which the system approaches its orthogonal state determines the metrological capability of the system. 
This relation views $\mathcal{F}_Q$ not as a merely variance-type measure of errors but also a fundamental speed measure, capturing the instantaneous separation rate of states along the parameter change. Consequently, the precision bound that follows is fundamentally connected to the quantum speed limit.
By viewing phase sensitivity as the geodesic speed in the Bures metric, the connection to the quantum Fisher information becomes explicit, allowing quantum speed limits to be recognized as metrological bounds rather than purely dynamical constraints.

\section{ Superresolving Rabi Oscillations}

We now adopt an effective \(m\)-photon atom-field interaction that captures super-resolving dynamics while remaining experimentally realistic. 
Operating at large detuning from single-photon resonance, we adopt an effective \(m\)-photon description with coupling \(\lambda\), which governs the dynamics and reaches experimentally realistic values in various systems. Therefore, we consider a nonlinear interaction Hamiltonian~\cite{villas2019multiphoton,felicetti2018two,zou2020multiphoton} of the form
\begin{equation}\label{Hamiltonian}
\mathcal{V} = \lambda \left( {a^{\dagger}}^m \sigma_- + a^m \sigma_+ \right),
\end{equation}
where $\lambda$ is the effective higher-order coupling coefficient, and $a^\dagger$ and $a$ are the photon creation and annihilation operators of the field mode. The operator $\sigma_-$ ($\sigma_+$) lowers (raises) the atomic state. 


The Hamiltonian in Eq.~\eqref{Hamiltonian} should be understood as an effective $m$-photon interaction Hamiltonian. It is valid when the system is tuned close to the desired $m$-photon resonance, while lower-order or unwanted transitions remain sufficiently off-resonant. In this regime, the intermediate states can be adiabatically eliminated and their effect is absorbed into the effective coupling constant $\lambda$, which depends on the microscopic coupling strengths and detunings. For example, in a two-photon process one typically obtains an effective coupling of the form $\lambda \sim g^2/\Delta$, where $g$ is the single-photon coupling and $\Delta$ is the detuning from the intermediate transition~\cite{toor1992validity}. Therefore, the model is not intended as a full microscopic description of a specific platform, but rather as an effective description that captures the dominant resonant multiphoton dynamics relevant for the metrological scaling. Such effective nonlinear couplings are experimentally motivated in strongly coupled light--matter systems, including cavity and circuit-QED platforms, provided that the effective coupling remains larger than the relevant decoherence and loss rates over the interrogation time. In particular, the single-photon coupling strength $g$ can be made sufficiently large in the strong-coupling regime, often reaching tens of MHz in experimental settings. This enables a relatively strong effective coupling $\lambda$, which can reach values on the order of hundreds of kHz~\cite{villas2019multiphoton,felicetti2018two,zou2020multiphoton}.

The Hamiltonian in Eq.~\eqref{Hamiltonian} has been shown to provide a valuable framework for various quantum information applications~\cite{villas2019multiphoton,felicetti2018two,zou2020multiphoton}. In this work, we demonstrate another important use of this Hamiltonian in quantum metrology. It is worth noting that a full microscopic treatment yields a slightly modified form of the Hamiltonian, introducing an additional phase factor in the two-photon process due to the dynamical Stark shift~\cite{toor1992validity}. Nevertheless, we adopt the effective Hamiltonian as presented above, as it offers a more transparent perspective for uncovering the key physical insights of the problem~\cite{villas2019multiphoton,felicetti2018two,zou2020multiphoton}.

We now consider a general initial state where the atom is in a superposition of states $|a\rangle$ and $|b\rangle$ as
\begin{equation}
|\psi(0)\rangle_{\text{atom}} = c_b |b\rangle + c_a |a\rangle,
\end{equation}
and the field is in the state
\begin{equation}
|\psi(0)\rangle_{\text{field}} = \sum_{n=0}^{\infty} c_n |n\rangle.
\end{equation}

The initial atom-field state is then
\begin{equation}
|\psi(0)\rangle = |\psi(0)\rangle_{\text{atom}} \otimes |\psi(0)\rangle_{\text{field}} = \sum_{n=0}^{\infty} \left[ c_{a,n} |a, n\rangle + c_{b,n} |b, n\rangle \right],
\end{equation}
where $c_{a,n} = c_a c_n$ and $c_{b,n} = c_b c_n$. Here, $|a,n\rangle$ is the state with the atom in its excited state $|a\rangle$ and the field having $n$ photons, and similarly, $|b,n\rangle$ is the state with the atom in its ground state $|b\rangle$ and the field having $n$ photons.

{Since we consider the interaction Hamiltonian, we define the slowly varying probability amplitudes $c_{a,n}(t)$ and $c_{b,n}(t)$, such that the state vector at time $t$ can be written as}
\begin{equation}
|\psi(t)\rangle = \sum_{n=0}^{\infty} \left[ c_{a,n}(t) |a, n\rangle + c_{b,n}(t) |b, n\rangle \right].
\end{equation}

The time evolution of the coefficients is governed by the general solution
\begin{align}
c_{a,n}(t) &= c_{a,n}(0) \cos\left( \frac{\Omega_{n,m} t}{2} \right)
- i\, c_{b,n+m}(0) \sin\left( \frac{\Omega_{n,m} t}{2} \right), \label{eq:ca} \\ \nonumber
c_{b,n+m}(t) &= c_{b,n+m}(0) \cos\left( \frac{\Omega_{n,m} t}{2} \right)
- i\, c_{a,n}(0) \sin\left( \frac{\Omega_{n,m} t}{2} \right), \label{eq:cb}
\end{align}
where the generalized Rabi frequency for the $m$-photon transition is given by
\begin{equation}
\Omega_{n,m} = 2\lambda \sqrt{ \frac{(n+m)!}{n!} }.
\end{equation}

If the atom is initially prepared in the ground state, then $c_{a,n}(0) = 0$ and $c_{b,n}(0) = c_n(0),$
where $c_n(0)$ denotes the initial probability amplitude of the field. In this case, the solutions simplify to
\begin{align}
c_{a,n-m}(t) &= -i c_n(0) \sin\left( \frac{\Omega_{n-m,m} t}{2} \right), \label{eq:ca_simplified} \\ \nonumber
c_{b,n}(t) &= c_n(0) \cos\left( \frac{\Omega_{n-m,m} t}{2} \right). \label{eq:cb_simplified}
\end{align}

For this scenario, the atomic population inversion is defined as~\cite{scully1999quantum}
\begin{equation}
W(t) = \sum_{n=0}^{\infty} \left[ |c_{a,n}(t)|^2 - |c_{b,n}(t)|^2 \right].
\end{equation}

The dynamics of atomic inversion is a fundamental observable in quantum optics. One of its most notable features is the collapse and revival phenomenon that arises when the field is initially in a coherent state~\cite{scully1999quantum}. To analyze the dynamics relevant to our setting, we consider the regime $n \gg m$, where the generalized Rabi frequency simplifies as $\Omega_{n,m} \approx \Omega_{n-m,m} \approx 2\lambda n^{m/2}$. Defining $\Omega(n) = 2\lambda n^{m/2}$, the atomic inversion can be approximated as
\begin{equation}
\begin{aligned}
    W(t) &\approx \sum_{n} |c_{n+m}(0)|^2 \sin^2\left( \frac{\Omega(n) t}{2} \right) - |c_n(0)|^2 \cos^2\left( \frac{\Omega(n) t}{2} \right) \\
         &= \sum_{n} |c_{n+m}(0)|^2 - \left[ |c_{n+m}(0)|^2 + |c_n(0)|^2 \right] \cos^2\left( \frac{\Omega(n) t}{2} \right).
\end{aligned}
\end{equation}

The oscillatory behavior is thus governed by the term $\cos^2\left( {\Omega(n) t}/{2} \right)$, which captures the dynamics of Rabi oscillations and dictates how quickly the system evolves toward an orthogonal state. As a result, the time resolution scales as $\delta t \propto 1/n^{m/2}$, where $n$ is the average photon number of the field. Therefore, for a fixed photon number  $n$ the scaling  can go beyond $1/n$
for higher order nonlinearities with $m>2$.

As a special case, consider the initial state of the system $|\psi(0)\rangle = |b, n\rangle$, where the atom is in the ground state and the field contains $n$ photons. The time evolution under the nonlinear interaction Hamiltonian leads to
\begin{align}\label{Fock}
|\psi(t)\rangle = \cos\left( \frac{\Omega_{n-m,m} t}{2} \right) |b, n\rangle - i \sin\left( \frac{\Omega_{n-m,m} t}{2} \right) |a, n - m\rangle.
\end{align}

\begin{figure}[h]
\includegraphics[width=\columnwidth]{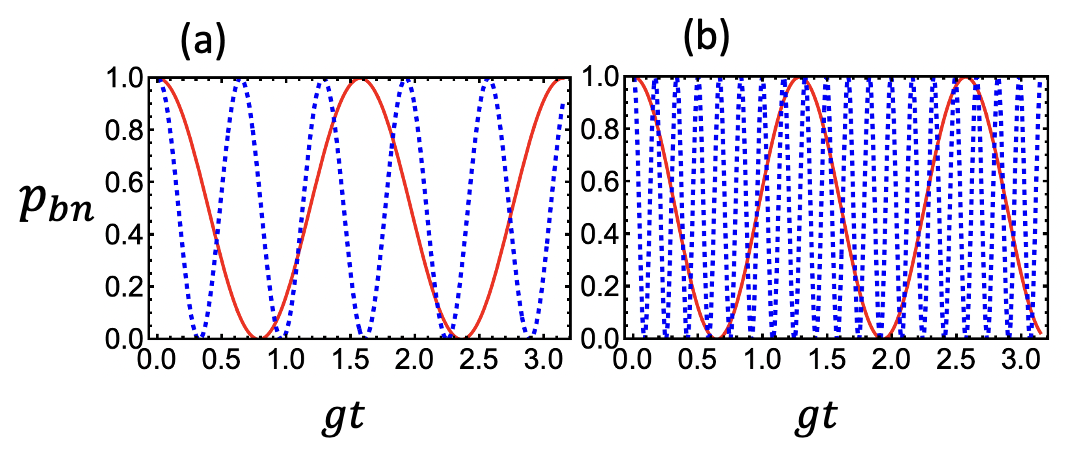}
\caption{
Probability of finding the atom in the ground state for the initial condition $|b,n\rangle$ with $n=4$ in (a) and $n=6$ in (b). The solid curves correspond to the single-photon process $(m=1)$, while the dotted curves correspond to the four-photon process $(m=4)$. The faster oscillations for larger $m$ show that higher-order multiphoton coupling drives the system through distinguishable states more rapidly.  This increased dynamical speed corresponds to a larger quantum Fisher information for time estimation and therefore supports the improved time-resolution scaling $\delta t \propto 1/\Omega_{n-m,m}$.
}
\label{fig:2}
\end{figure}

The probability of finding the atom in the ground state at time $t$ is given by 
\begin{equation}
p_{bn}(t) = \cos^2\left( \frac{\Omega_{n-m,m} t}{2} \right), \end{equation}
which is plotted in Fig.~\ref{fig:2}. The oscillation frequency, and thus the resolution, is governed by the effective coupling factor $k = \sqrt{n! / (n - m)!}$, which increases with both photon number $n$ and photon order $m$.

In Fig.~\ref{fig:2}(a), we consider the case $n = 4$. The solid (red) curve corresponds to $m = 1$, yielding $k = 2$, while the dotted (blue) curve corresponds to $m = 4$, with $k \approx 5$. As expected, the oscillation frequency for $m = 4$ is approximately 2.5 times higher than that for $m = 1$ over the same time interval. In Fig.~\ref{fig:2}(b), we examine the case $n = 6$, again comparing $m = 1$ and $m = 4$, which correspond to $k = \sqrt{6}$ and $k = \sqrt{360}$, respectively.

These results clearly illustrate that even a modest increase in the photon number $n$ leads to a substantial enhancement in the resolution of nonlinear Rabi oscillations. This behavior underscores the potential of such dynamics to function as a superresolving quantum clock, with time resolution scaling as $\delta t \propto 1/\Omega_{n-m,m}$. Furthermore, when employing $R$ independent atoms, the precision can be improved in accordance with the quantum Cram\'er--Rao bound, yielding 
\begin{equation}
\delta t_{\mathrm{CRB}} \propto \frac{1}{\sqrt{R} \Omega_{n-m,m}}.    
\end{equation}

Using $\Omega_{n-m,m}\simeq 2\lambda\,n^{m/2}$ for $n\gg m$, this equation yields the explicit bound
\begin{align}
    \delta t_{\mathrm{CRB}} \approx \frac{1}{2\sqrt{R}\,\lambda\,n^{m/2}}\,.
\end{align}

This expression shows that the time resolution improves polynomially with photon number with power $m/2$, and gains the usual $1/\sqrt{R}$ enhancement from $R$ independent repetitions. 

\section{ Generation of the ON State}

The simple configuration described in Eq.~\eqref{Fock} can lead to an interesting scenario. For ${\Omega_{n-m,m} t} = \pi/2$, the system evolves into a specific entangled state between the atom and the field,
\begin{align}
    |\psi\rangle = \frac{1}{\sqrt{2}}\left( |b, n\rangle - i |a, n - m\rangle \right).
\end{align}

If the atom is subsequently measured in the superposition state $(|b\rangle - i |a\rangle)/\sqrt{2}$, the field collapses into a superposition of two Fock states, $|n\rangle$ and $|n - m\rangle$, given by
\begin{align}
    |\psi\rangle = \frac{1}{\sqrt{2}}\left( |n\rangle + |n - m\rangle \right).
\end{align}

As a particular case, when $n = m$, the resulting field state becomes a superposition of the vacuum and the $n$-photon Fock state, written as
\begin{align}
    |\psi\rangle = \frac{1}{\sqrt{2}}\left( |0\rangle + |n\rangle \right).
\end{align}

This state serves as the single-mode analogue of the NOON state and we refer to here as the ON state.

We now examine the metrological performance of the ON state. A phase shift can be imprinted on the state via the unitary operator $e^{-i \varphi \hat{n}}$, yielding
\begin{align}
    |\psi(\varphi)\rangle = e^{-i \varphi \hat{n}} |\psi(0)\rangle = \frac{1}{\sqrt{2}} \left( |0\rangle + e^{-i \varphi n} |n\rangle \right).
\end{align}

To estimate the phase $\varphi$, we introduce the observable $P = |0\rangle\langle n| + |n\rangle\langle 0|$ that is used for the extraction of the phase information. 

Therefore, the expectation value of this observable with respect to the state $|\psi(\varphi)\rangle$ is
\begin{align}
    \langle P \rangle = \langle \psi(\varphi) | P | \psi(\varphi) \rangle = \cos(n \varphi),
\end{align}
and its standard deviation is given by $\Delta P = \sin(n \varphi)$. Using the error propagation formula, the phase uncertainty is obtained as
\begin{align}
    \delta \varphi = \frac{\Delta P}{|d\langle P\rangle / d\varphi|} = \frac{1}{n}.
\end{align}

Since the average photon number in the state is $\langle \hat{n} \rangle = n/2$, this result corresponds to the Heisenberg limit of phase precision. What is particularly noteworthy here is that this $1/n$ scaling is often associated with the NOON state, where it arises due to entanglement between two modes. However, the ON state achieves the same Heisenberg-limited sensitivity without requiring any mode entanglement. This demonstrates that single-mode superpositions can also attain optimal precision in quantum metrology.

Interestingly, this state has been discussed previously in the literature [see Ref. ~\cite{zwierz2010general}], where it was noted that such a superposition can in principle achieve Heisenberg-limited scaling, although no clear method for its realization was introduced. The formalism developed here provides a concrete mechanism for generating this state and may serve as a basis for experimentally demonstrating Heisenberg-limited scaling without the need for entanglement, using the ON state. 

\section{ From Quantum Speed Limit to Ultimate Metrological Scaling}

To gain a deeper understanding of the sensitivity of Rabi oscillations, it is instructive to analyze the system from the perspective of QFI. For a pure state $|\Psi(t)\rangle$, the QFI with respect to time is given by Eq.~\eqref{Fisher}. Applying this expression to the state in Eq.~\eqref{Fock} yields $\mathcal{F}_Q(t) = \Omega_{n-m,m}^2$. This directly implies that the error in time estimation scales as $\delta t \propto 1/{\Omega_{n-m,m}}$, or equivalently,  
\begin{equation}
\delta t \propto \left[ \frac{(n-m)!}{n!} \right]^{1/2},    
\end{equation}
which asymptotically approaches the scaling $\delta t_{\mathrm{CRB}} \propto 1/n^{m/2}$ for large $n$. Using the expression for the instantaneous speed of quantum evolution in the Bures geometry, $\mathcal{V}(t) = \frac{dS_B}{dt} = \frac{1}{2} \sqrt{\mathcal{F}_Q(t)},$
we find
\begin{equation}
\mathcal{V}(t) = \frac{\Omega_{n-m,m}}{2}.
\end{equation}

This indicates that the speed of evolution increases with both the photon number $n$ and the nonlinearity order $m$.

More generally, the QFI for an arbitrary mixed state $\rho$ is bounded by
\begin{equation}
\mathcal{F}_Q(t) \leq \frac{4 (\Delta \mathcal{H})^2}{\hbar^2},    
\end{equation}
where $\Delta \mathcal{H}$ is the standard deviation (energy uncertainty) of the system Hamiltonian $\mathcal{H}$.

The first formulation of the QSL was introduced by Mandelstam and Tamm, who demonstrated that the minimum time required for a quantum system to evolve between two orthogonal states is bounded from below by ~\cite{mandelstam1991uncertainty}
\begin{align}
    t \geq \frac{\pi \hbar}{2 \Delta \mathcal{H}}.
\end{align}

This bound was later generalized to arbitrary (nonorthogonal) quantum states $\rho$ and $\sigma$ through the introduction of the Bures angle $\mathcal{L}(\rho, \sigma)$, leading to~\cite{deffner2017quantum}

\begin{align}
    t \geq \frac{\hbar}{\Delta \mathcal{H}}\, \mathcal{L}(\rho, \sigma).
\end{align}

In the limit of infinitesimal time evolution, where $\rho(t)$ evolves continuously to $\rho(t + \delta t)$, the Bures angle reduces to $\mathcal{L}(\rho(t), \rho(t + \delta t)) = dS_B$. This yields the inequality
\[
\mathcal{V}(t) = \frac{1}{2} \sqrt{\mathcal{F}_Q(t)} \leq \frac{\Delta \mathcal{H}}{\hbar}.
\]

This bound establishes a fundamental connection between operator variance and the concept of QSL~\cite{maleki2023speed}.

Therefore, the minimum resolvable time interval is given by

\begin{align}
    \delta t_{\mathrm{CRB}} = \frac{1}{\sqrt{\mathcal{F}_Q(t)}} \geq \frac{\hbar}{2 \Delta \mathcal{H}}.
\end{align}

This relation demonstrates that the QSL imposes a fundamental constraint on the attainable precision of quantum metrological protocols, directly linking the energy fluctuations of the probe to its ultimate estimation capability.

The $1/n$ Heisenberg limit associated with the NOON state is widely recognized as the optimal scaling for quantum metrology \cite{kim2025distributed}. However, as demonstrated above, the precision scaling with $n$ photons at the input can instead follow $\delta t_{\mathrm{CRB}} \propto 1/n^{m/2}$, enabling sensitivities that surpass the conventional $1/n$ scaling of the NOON state, as illustrated in Fig.~\ref{fig:2}. At first glance, this appears to challenge the notion that the NOON state achieves the ultimate metrological precision. Such observations are not confined to the present work. 
For example, Beltr{\'a}n and Luis showed that phase estimation in systems with classical optical nonlinearities can exhibit an error scaling of $n^{-3/2}$~\cite{beltran2005breaking}, a result that was later confirmed experimentally by Napolitano \emph{et al.}~\cite{napolitano2011interaction}. Even stronger scalings of the form $n^{-q}$, where $q$ is a positive integer, have been reported by Boixo \emph{et al.}~\cite{boixo2007generalized}. These findings suggest that, under certain nonlinear or interaction-based strategies, one can achieve precision beyond the traditional Heisenberg limit. This effect is often referred to as "super-Heisenberg" scaling~\cite{napolitano2011interaction,rams2018limits}. Consequently, the conventional Heisenberg limit cannot be regarded as a universally valid bound on metrological precision. This raises a fundamental question: what is the true quantum-mechanical limit that dictates the ultimate scaling of parameter estimation?

To resolve the apparent paradox surrounding the optimal scaling of precision, it is essential to adopt a consistent and physically meaningful definition of resource counting. As previously discussed, the QFI is bounded by the variance (uncertainty) of the system Hamiltonian, linking metrological precision to the Mandelstam-Tamm bound. In certain specific scenarios, this variance leads to a number-phase uncertainty relation, which underpins the conventional usage of the term Heisenberg limit~\cite{zwierz2010general,holland1993interferometric}. This connection suggests that the metrological resource should be identified with the generator of the parameter shift, typically the system Hamiltonian.
However, using the Hamiltonian variance as a universal measure of resource consumption is problematic. In particular, it does not always provide a well-defined or finite quantity. For example, in certain physically relevant systems, the variance of the Hamiltonian may diverge. A prominent case is the class of systems characterized by a Breit-Wigner energy spectrum~\cite{breit1936capture,uffink1993rate}, where the energy uncertainty becomes ill-defined. This highlights the need for a more robust and universally applicable framework for defining resources in quantum metrology.

It turns out that framing the problem within the context of the QSL provides an alternative way to bound the achievable precision in estimating a phase shift (or equivalently, time). An important result in this regard is the bound introduced by Margolus and Levitin, which states that the minimum time required for a quantum system to evolve into an orthogonal state is $t \geq \frac{\pi \hbar}{2 \langle \mathcal{H} \rangle}$~\cite{margolus1998maximum, kominis2024quantum},
where $\langle \mathcal{H} \rangle$ denotes the average energy of the system, whereas $\langle \hat n\rangle$ denotes only the average number of photons in the field.
These two quantities coincide only for simple linear phase shifts generated by $\hat n$. 
In nonlinear or multiphoton protocols, however, the generator can depend nonlinearly on photon number, for example through $\hat n^m$, so the relevant resource is not simply $\langle \hat n\rangle$, but rather the expectation value of the generator, such as $\langle \hat n^m\rangle$.

When $\mathcal{H}$ represents a physical Hamiltonian, the absolute energy origin carries no operational meaning; adding a constant leaves the dynamics unchanged but shifts $\langle \mathcal{H} \rangle$. We therefore fix the reference by assigning zero energy to the ground state (minimum energy state). This can be done by using the shift
\begin{equation}
\mathcal{H}'=\mathcal{H}-h_{\min} I,
\end{equation}
where $h_{\min}$ is the smallest eigenvalue and $I$ is the identity.
Resource measures and precision bounds that depend on the mean energy should be expressed in terms of $\langle \mathcal{H}' \rangle$, the average energy measured from the ground state. This quantity is invariant under energy shifts and aligns with the form required by Margolus-Levitin bounds.

Similar to the Mandelstam-Tamm bound, this result has been generalized to arbitrary (nonorthogonal) quantum states using the Bures angle $\mathcal{L}(\rho_0, \rho_T)$, leading to the inequality~\cite{campaioli2018tightening}
\begin{align}
    t \geq \frac{\hbar}{\langle \mathcal{H} \rangle} \, \mathcal{L}(\rho_0, \rho_T).
\end{align}
From this relation, one sees that the Margolus-Levitin bound implies a fundamental constraint on estimation precision, $\delta\varphi_{\mathrm{CRB}} \propto 1/\langle \mathcal{H} \rangle$. Therefore, any quantum metrological strategy must respect this bound in addition to the one set by the energy variance.
Crucially, unlike the variance, the average energy $\langle \mathcal{H} \rangle$ is always positive and finite, making it a well-defined and operationally meaningful measure of the resource. Intuitively, increasing the average energy allocated to the system enables higher precision, while $\langle \mathcal{H} \rangle = 0$ implies no informational gain, as no physical resource is available to encode or extract information.

Therefore, for the nonlinear interaction Hamiltonian in Eq.~\eqref{Hamiltonian}, the appropriate definition of the resource is given by $\mathcal{H} = (a^\dag a)^m$, reflecting the $m$-photon nature of the coupling. As a result, the resource count is characterized by the average value $\langle \mathcal{H} \rangle = \langle n^m \rangle$.  Strategies based on classical pulses typically yield a precision scaling of $1/\langle n \rangle^{m/2}$, which corresponds to $1/\langle \mathcal{H} \rangle^{1/2}$. The Rabi oscillation protocol developed here surpasses the conventional $1/n$ scaling, highlighting its potential as a practically relevant approach for achieving super-Heisenberg precision in quantum metrology.

Therefore, when the phase-generating operator is defined as $\mathcal{H} = (a^\dag a)^m$, its application to a NOON state enables phase estimation under nonlinear dynamics. The phase shift is imprinted on the state as  
\begin{align}
    |\psi(\varphi)\rangle = \exp(-i \varphi \hat{n}^m) |\psi(0)\rangle = \frac{1}{\sqrt{2}} \left( |n0\rangle + e^{-i \varphi n^m} |0n\rangle \right).
\end{align}

To extract the phase information, we introduce the observable $P = |n0\rangle\langle 0n| + |0n\rangle\langle n0|$. The expectation value of this observable with respect to the evolved state is  
\begin{align}
    \langle P \rangle = \langle \psi(\varphi) | P | \psi(\varphi) \rangle = \cos(n^m \varphi).
\end{align}
This gives $\Delta P=\sin(n^m \varphi)$,
from which the phase uncertainty can be derived using error propagation
\begin{align}
    \delta\varphi = \frac{\Delta P}{|d\langle P \rangle / d\varphi|} = \frac{1}{n^m}.
\end{align}

This result saturates the fundamental scaling limit $\delta \varphi \propto 1/\langle \mathcal{H} \rangle$ imposed by the Margolus-Levitin bound. In the case of $m = 1$, this reduces to the well-known Heisenberg scaling $\delta\varphi = 1/n$, confirming that the NOON state achieves optimal precision under both linear and nonlinear phase shift generators.

As a result, the apparent paradox in precision scaling can be resolved by identifying $\langle \mathcal{H} \rangle$ as the true resource of the system, leading to a Heisenberg-limit precision that scales as $1/\langle \mathcal{H} \rangle$, rather than $1/n$. Consequently, the Heisenberg limit should be understood in the context of the Margolus-Levitin bound rather than the traditional Heisenberg uncertainty relation. The latter is instead associated with the Mandelstam-Tamm bound, which relates the attainable precision to the variance of the Hamiltonian $\mathcal{H}$. As demonstrated above, our parameter estimation framework simultaneously satisfies both bounds. Therefore, it becomes evident that many of the so-called super-Heisenberg scaling protocols, which appear to surpass the conventional Heisenberg limit, in fact do not reach the ultimate quantum bound on precision.

It is important to note that the scaling laws derived above correspond to an ideal coherent evolution. In realistic implementations, decoherence, photon loss, imperfect state preparation, finite detection efficiency, and fluctuations in the effective coupling strength can reduce the achievable Fisher information \cite{chin2012quantum,zhou2023emission,shaji2007qubit}. These effects are especially relevant for higher-order multiphoton processes, where the effective coupling $\lambda$ is typically weaker and the required coherence time may become more demanding. Therefore, the condition for observing the predicted scaling is not only that the effective $m$-photon Hamiltonian is valid, but also that the coherent Rabi dynamics occurs on a timescale shorter than the relevant decoherence and loss times.

For the ON-state protocol, losses and dephasing are also important because the metrological advantage depends on preserving coherence between the vacuum and the $n$-photon component. Photon loss can partially reveal which-number information and suppress the off-diagonal coherence responsible for the phase-dependent signal $\langle P\rangle=\cos(n\phi)$. Similarly, detector inefficiency and imperfect projection measurements reduce the visibility of the measured oscillations and therefore lower the attainable precision \cite{datta2011quantum}. Thus, although the ON state shows that $1/n$ scaling does not by itself certify mode entanglement as the enabling resource, its practical realization still requires high-fidelity state preparation, low-loss evolution, and efficient readout.

These considerations do not change the resource-counting argument of the present work. Rather, they clarify that the results should be interpreted as ideal quantum speed-limit and Fisher-information bounds. In experiments, decoherence and imperfections will generally reduce the prefactor and may limit the range over which the asymptotic scaling can be observed. Nevertheless, platforms with strong light--matter coupling, long coherence times, and controlled multiphoton interactions provide a realistic route for testing the predicted dynamics and for exploring the resource-consistent interpretation of apparent super-Heisenberg scaling.

\section{Conclusions}

We presented a resource-consistent framework that ties metrological precision directly to quantum dynamical speed, clarifying when and how super-Heisenberg-like scalings arise. By analyzing multiphoton Rabi dynamics, we showed that the time (or phase) uncertainty can scale as $\delta t \propto n^{-m/2}$ with average photon number $n$ and nonlinearity order $m$, and we identified the correct resource for such protocols as the generator's expectation $\langle H\rangle$ rather than photon number alone. Framing precision through quantum speed limits reconciles these scalings with fundamental bounds: the Mandelstam--Tamm relation links precision to energy variance, while the Margolus--Levitin bound shows that the ultimate Heisenberg regime is $1/\langle H\rangle$. Within this unified view, seemingly paradoxical beyond-$1/n$ behaviors reflect different generators and resource accounting, not a violation of Heisenberg scalings. In other words, the Heisenberg limit
should be understood in the context of the Margolus-Levitin type speed limit
bounds rather than the traditional Heisenberg uncertainty relation.
Moreover, we revisited the common attribution of the NOON state's $1/n$ scaling to entanglement. We considered a setting showing that the $1/n$ scaling alone does not certify entanglement as the enabling resource in this setting. We explicitly construct the feasible single-mode ON state probe that achieves the same $1/n$ sensitivity without mode entanglement.

These findings illuminate the ultimate scaling of quantum sensors and  clarify the proper interpretation of the so-called Heisenberg scalings. These results shed light on the connection between quantum metrology and quantum speed limits, and show that demonstrating genuine quantum advantages requires careful, system-specific quantification of the operative resources.

\section*{Acknowledgements}

I would like to thank Marlan O. Scully and M. Suhail Zubairy for insightful discussions. 



\nocite{apsrev41Control}
\bibliographystyle{apsrev4-1}
\bibliography{Refs}

\end{document}